\documentclass[english]{article}
\usepackage[preprint]{waspaa19}
\usepackage{todonotes}
\usepackage{amsmath,amssymb,gensymb}
\usepackage{graphicx}
\usepackage{cite}

\begin{document}

\copyrightnotice{\copyright2019 IEEE}
\toappear{To appear at WASPAA 2019}

\title{MOTION-TOLERANT BEAMFORMING WITH DEFORMABLE MICROPHONE ARRAYS}

\name{Ryan M. Corey and Andrew C. Singer\thanks{\hspace{-5mm}This material is based upon work supported by the National Science Foundation Graduate Research Fellowship Program under grant no. DGE-1144245.}}

\address{University of Illinois at Urbana-Champaign}

\ninept
\maketitle

\begin{abstract}
Microphone arrays are usually assumed to have rigid geometries:
the microphones may move with respect to the sound field but remain
fixed relative to each other. However, many useful arrays, such
as those in wearable devices, have sensors that can move relative to each other. We compare two
approaches to beamforming with deformable microphone arrays: first,
by explicitly tracking the geometry of the array as it changes over
time, and second, by designing a time-invariant beamformer based on the second-order statistics of the moving array.
The time-invariant approach is shown to be appropriate
when the motion of the array is small relative to the acoustic wavelengths
of interest. The performance of the proposed beamforming system
is demonstrated using a wearable microphone array on a moving human
listener in a cocktail-party scenario.
\end{abstract}


\section{Introduction}

Microphone arrays can be used to spatially localize and separate sound sources
from different directions \cite{Benesty2008,gannot2017consolidated,makino2018audio,vincent2018audio}. Small arrays,
typically with up to eight microphones spaced a few centimeters
apart, are widely used in teleconferencing and speech recognition. A promising application is in hearing
aids and other augmented listening devices \cite{doclo2015magazine},
where arrays could improve intelligibility in noisy environments.
However, the arrays in listening devices are tiny: typically
only two microphones a few millimeters apart. 

\begin{figure}
\begin{centering}
\includegraphics[height=5cm]{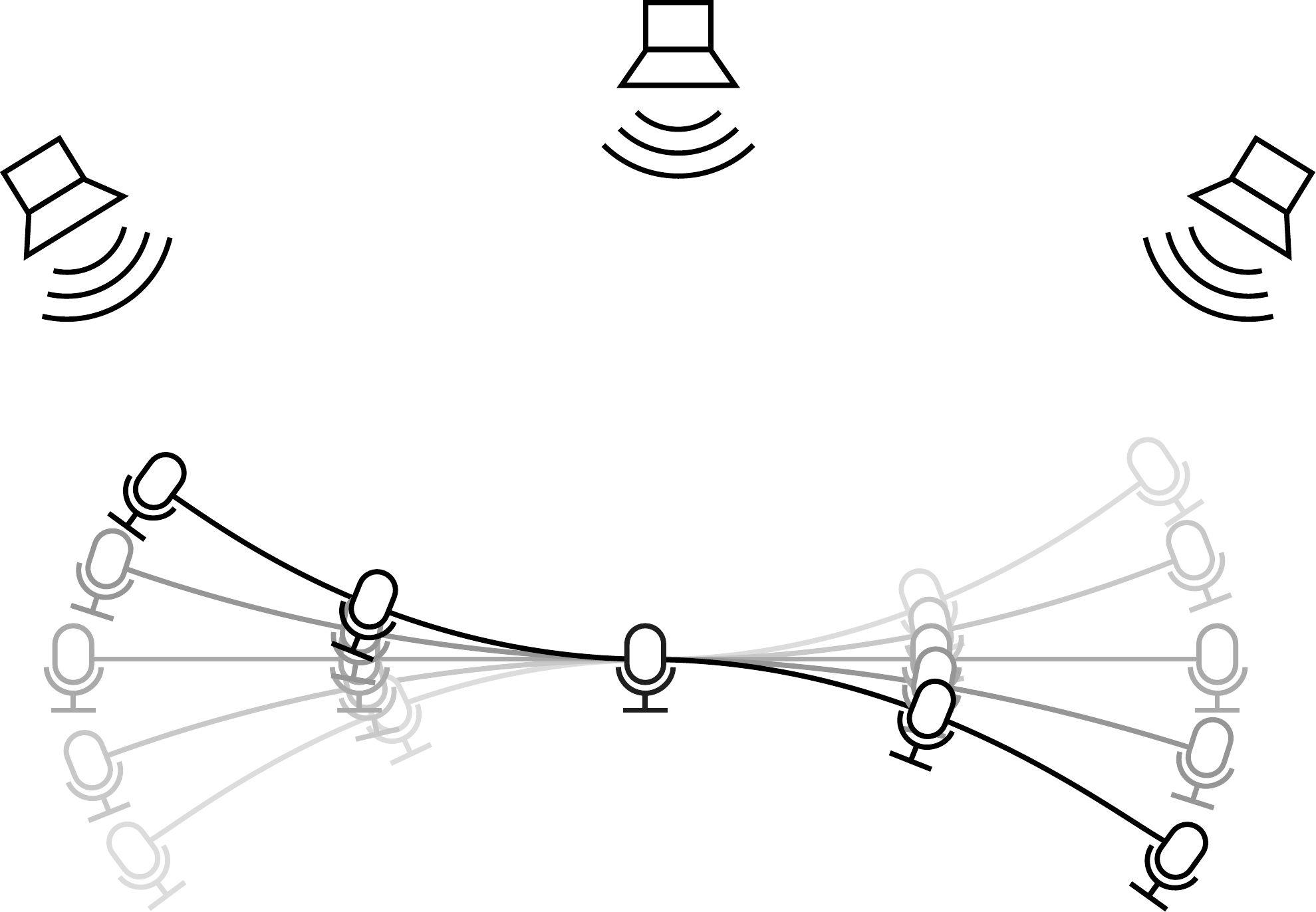}
\par\end{centering}
\caption{\label{fig:motivation}In a deformable microphone array, the sensors
can move relative to the sound sources and also relative to each other.}
\end{figure}

Arrays with microphones spread across the body
can can perform better than listening devices
with only a few microphones near the ears \cite{corey2019brtf}. There
is a major challenge in using such arrays, however: humans move. The
microphones in a wearable array not only move relative to sound sources,
{\em but also move relative to each other}, as shown in Figure \ref{fig:motivation}.
Because array processing typically relies on phase differences between sensors, even small deformations can harm the performance
of a spatial sound capture system. 

There has been little prior work on deformable microphone arrays. In \cite{barfuss2014adaptive}, a robot with 
microphones on movable arms was adaptively repositioned to improve beamforming performance.
In \cite{bando2013posture}, microphones were placed along a hose-shaped
robot and used to estimate its posture. In \cite{corey2018async},
wearable arrays were placed on three human listeners in a cocktail
party scenario and aggregated using a sparsity-based time-varying
filter. That paper applied the full-rank covariance model for deformation that is presented here. 

In contrast, the problem of tracking moving \emph{sources} has received significant
attention. Most solutions combine
a localization method, such as steered response power or multiple
signal classification, with a tracking algorithm, such as Kalman or
particle filtering \cite{vermaak2001nonlinear,ward2003particle,valin2007robust,traa2014ransac,kounades2016variational,nikunen2018separation}.
Others use blind source separation techniques that adapt over time
as the sources move \cite{mukai2003robust,malek2012semi}. Sparse
signal models can improve performance when there are multiple competing
sound sources \cite{roman2008binaural,zhong2009time,golan2010subspace,higuchi2014underdetermined,corey2018async}.
These time-varying methods
are necessary when the motion of the sources or microphones is
large. However, tracking algorithms are computationally complex
and time-varying filters can introduce disturbing artifacts. For small
motion, such as breathing or nodding with a wearable array,
it may be possible to account for motion using a linear time-invariant
filter instead. 

The design of spatial filters that are robust to small
perturbations is well studied. Mismatch between the true
and assumed positions of the sensors can be modeled as uncorrelated 
noise and addressed using diagonal loading on the noise covariance matrix or using a norm constraint on the beamformer coefficient vector \cite{cox1987robust}. 
Other approaches include derivative constraints that
ensure the beam pattern does not change too quickly \cite{er1983derivative} and distortion constraints 
within a region or subspace \cite{zheng2004robust}.
For far-field beamformers, these methods widen
the beam pattern and therefore reduce array gain compared to
non-robust beamformers.

In this work, we explore the impact of deformation on the performance
of multimicrophone audio enhancement systems. If motion is small enough that it can
be effectively modeled using second-order statistics, then the signals
can be separated using linear time-invariant filters. Larger motion
destroys the spatial correlation structure of the sources and therefore
requires more complex time-varying methods. We compare the performance
of different beamforming strategies on two deformable arrays:
a linear array of microphones hanging from a pole, the motion of
which is straightforward to model, and a wearable array on a human
listener with more complex movement patterns. We find that the effects
of deformation are dramatic at high frequencies but manageable at
the low frequencies for which large arrays have the greatest benefit.

\section{Time-Frequency Beamforming}

Let $\mathbf{X}[t,f]=\left[X_{1}[t,f],X_{2}[t,f],\dots,X_{M}[t,f]\right]^{T}$
be the vector of short-time Fourier transforms (STFT) of the signals
captured at microphones $1$ through $M$, where $t$ is a time index
and $f$ is a frequency index. Assuming linear mixing, the received
signal can be modeled as the sum of components $\mathbf{C}_{1}[t,f],\dots,\mathbf{C}_{N}[t,f]$
due to $N$ sources and diffuse additive noise $\mathbf{V}[t,f]$:
\begin{equation}
\mathbf{X}[t,f]=\sum_{n=1}^{N}\mathbf{C}_{n}[t,f]+\mathbf{V}[t,f].
\end{equation}
The components $\mathbf{C}_{1},\dots,\mathbf{C}_{N}$ are sometimes
called source spatial images \cite{vincent2006performance}. Assume
that the source images and noise are zero-mean random processes that
are uncorrelated with each other and that the diffuse noise is wide-sense
stationary. Let $\mathbf{R}_{n}[t,f]=\mathbb{E}\left[\mathbf{C}_{n}[t,f]\mathbf{C}_{n}^{H}[t,f]\right]$
be the time-varying STFT covariance matrix of source image $n$ for
$n=1,\dots,N$, where $\mathbb{E}$ denotes expectation, and let $\mathbf{R}_{v}[f]$
be the time-invariant covariance of $\mathbf{V}[t,f]$.

The output $\mathbf{Y}[t,f]=\mathbf{W}[t,f]\mathbf{X}[t,f]$ of the
audio enhancement system is a linear transformation of the microphone
input signals in the time-frequency domain. The beamforming weights
$\mathbf{W}[t,f]$ may vary over time and may produce one or several outputs. In this work,
we restrict our attention to the multichannel Wiener filter (MWF) \cite{gannot2017consolidated},
which minimizes mean squared error between the output and a desired signal
$\mathbf{D}[t,f]$:
\begin{equation}
\mathbf{W}[t,f]=\mathrm{Cov}\left(\mathbf{D}[t,f],\mathbf{X}[t,f]\right)\mathrm{Cov}\left(\mathbf{X}[t,f]\right)^{-1}.
\end{equation}
Here we choose $\mathbf{D}[t,f]=\left[\mathbf{e}_1^{T}\mathbf{C}_{1}[t,f],\dots,\mathbf{e}_1^{T}\mathbf{C}_{N}[t,f]\right]^T$ where $\mathbf{e}_1 = [1,0,\dots,0]^T$;
that is, we estimate each source signal as observed at microphone
1. In a listening device, this reference microphone might be the one
nearest the ear canal so that head-related acoustic effects are preserved
\cite{doclo2006theoretical}. The MWF beamforming weights are given
by
\begin{align}
\mathbf{W}[t,f] & =\left[\begin{matrix}\mathbf{e}_1^{T}\mathbf{R}_{1}[t,f]\\
\vdots\\
\mathbf{e}_1^{T}\mathbf{R}_{N}[t,f]
\end{matrix}\right]\left(\sum_{n=1}^{N}\mathbf{R}_{n}[t,f]+\mathbf{R}_{v}[f]\right)^{-1}.\label{eq:MWF}
\end{align}

\subsection{Statistical models}

Many audio source separation and enhancement methods \cite{makino2018audio,vincent2018audio}
use time-varying STFT beamformers similar to (\ref{eq:MWF}).
Time-varying covariance matrices capture the nonstationarity of natural signals such as speech and adapt to source and microphone movement. Because the focus of this paper is on the spatial separability of
sound sources with deformable arrays, we will ignore the temporal
statistics of the sound sources. Any variation of $\mathbf{R}_{n}[t,f]$
with respect to $t$ is assumed to be due to motion of the microphones.

Let $\tilde{\mathbf{R}}_{n}[f;\theta]$ be the source covariance matrix
corresponding to state $\theta\in\mathcal{X}$ for $n=1,\dots,N$,
where $\mathcal{X}$ is a set of states that represent the positions
and orientations of the microphones. Assume that the motion of the
array is slow enough that each frame has a single corresponding state
$\Theta[t]$ and that the effects of Doppler can be neglected. Then the
sequence of covariance matrices is $\mathbf{R}_{n}[t,f] =\tilde{\mathbf{R}}_{n}[f;\Theta[t]]$ for  $n=1,\dots,N$.

While it is often assumed that each $\mathbf{R}_{n}$ is a rank-one matrix proportional
to the outer product of a steering vector, here we adopt the full-rank
STFT covariance model \cite{duong2010fullrank}. Although originally
developed to compensate for long impulse responses, the full-rank
model is also useful for modeling uncertainty due to deformation.

\begin{figure}
\begin{centering}
\includegraphics[height=3.5cm]{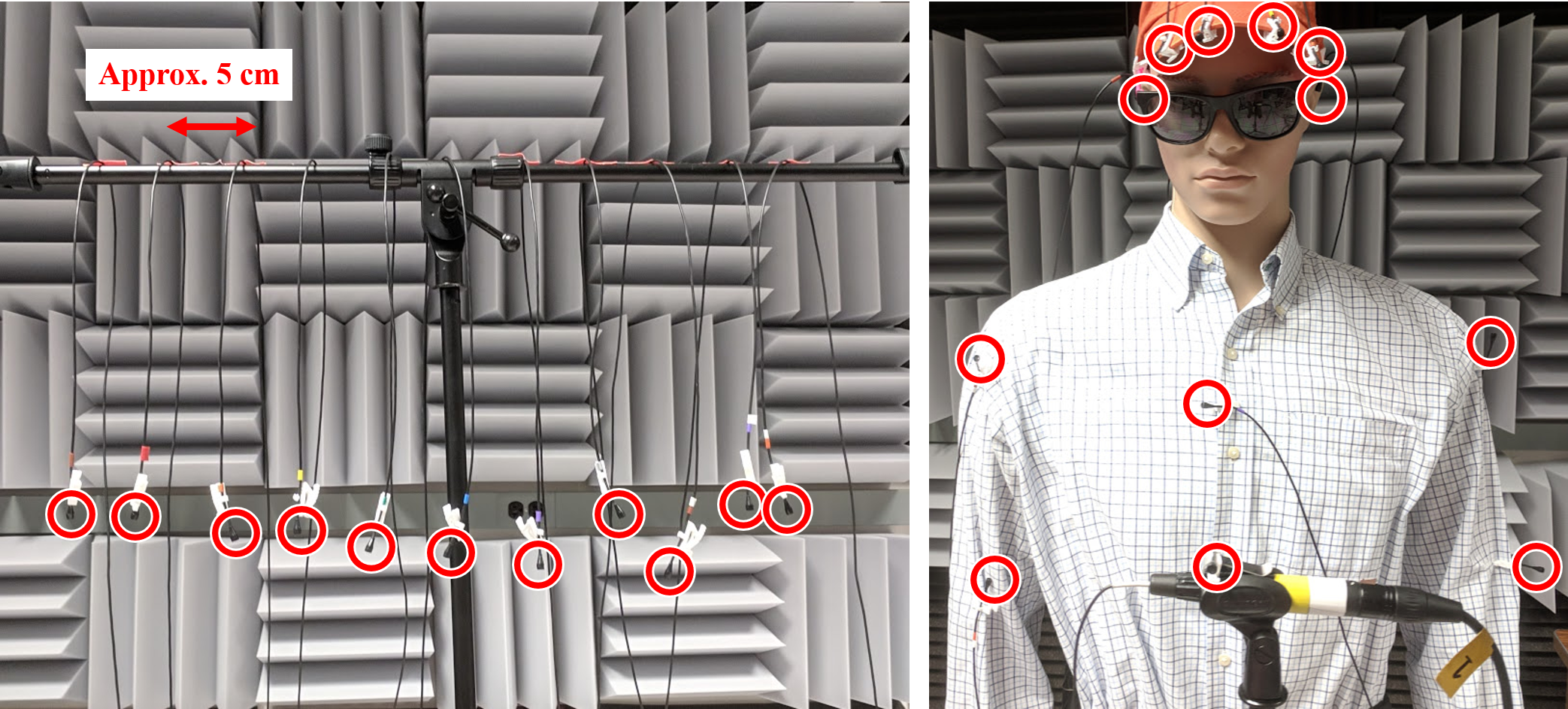}
\par\end{centering}
\caption{\label{fig:setup}Deformable linear array (left) and wearable array (right).}
\end{figure}

\subsection{Static and dynamic beamformers}

This work will compare the performance of two separation methods,
one static and one dynamic. For the static method, assume a prior
distribution $p_{\theta}$ on $\theta$. Because $\mathbf{C}_{n}[t,f]$
is assumed to have zero mean, the ensemble covariance matrices $\bar{\mathbf{R}}_{n}[f]$
are given by 
\begin{equation}
\bar{\mathbf{R}}_{n}[f]=\mathbb{E}\left[\mathrm{Cov}\left(\mathbf{C}_{n}\mid\Theta\right)\right]=\int_{\mathcal{\mathcal{X}}}p_{\theta}(\theta)\mathbf{\tilde{R}}_{n}[f;\theta]\,\mathrm{d}\theta,
\end{equation}
for $n=1,\dots,N$. The static beamformer is computed by substituting
$\bar{\mathbf{R}}_{n}[f]$ for $\mathbf{R}_{n}[t,f]$ in (\ref{eq:MWF}).
In the static beamforming experiments presented here, the states 
are never explicitly defined. Instead, each $\bar{\mathbf{R}}_{n}[f]$
is estimated by the sample covariance over a set of training data. This is equivalent to an empirical measure over $\Theta$.

For the dynamic method, assume that an estimate $\hat{\Theta}[t]$
of the state sequence is available, for example from a tracking algorithm.
Then the estimated covariance matrices are 
\begin{equation}
\hat{\mathbf{R}}_{n}[t,f]=\tilde{\mathbf{R}}_{n}[f;\hat{\Theta}[t]],\quad n=1,\dots,N.
\end{equation}
In the results presented here, the set of states is manually determined
for each experiment based on the range of motion of the array. For
example, the linear array has discrete states representing different angles
of rotation. To ensure that the results are as general as possible,
we do not use a blind state estimation or tracking algorithm. Instead,
we measure the states using near-ultrasonic pilot signals
that are played back alongside the source speech signals. The source statistics within each discrete state are estimated by the sample covariance of the training data for time frames in that state.

\section{Second-Order Statistics}

Because the MWF depends on the second-order
statistics of the observed signals, it will be instructive to analyze
the effects of deformation on the covariance structure of the acoustic
source images. 

Since the source images are assumed to have full rank, they do not occupy different
subspaces and the separability of
different sources must be analyzed statistically. For example, the Kullback-Leibler
divergence between two zero-mean multivariate Gaussian distributions
with covariances $\mathbf{R}_{1}$ and $\mathbf{R}_{2}$ is \cite{levy2008principles}
\begin{equation}
D(\mathbf{R}_{1},\mathbf{R}_{2})=\frac{1}{2}\left[\mathrm{trace}\left(\mathbf{R}_{1}\mathbf{R}_{2}^{-1}-\boldsymbol{I}\right)-\ln\frac{\mathrm{det}\mathbf{R}_{1}}{\mathrm{det}\mathbf{R}_{2}}\right].\label{eq:div_gaussian}
\end{equation}
This quantity is largest for pairs of matrices whose principal eigenvectors
are orthogonal and zero for identical
matrices. Although the signals captured by deformable arrays do
not have Gaussian distributions, the divergence expression
(\ref{eq:div_gaussian}) will be useful in quantifying the impact
of deformation on their second-order statistics.

\subsection{Ideal far-field array}

Consider an array of ideal isotropic sensors observing $N$ far-field
sources from different angles. Suppose that the sources
all have power spectral density $\sigma_{n}^{2}[f]=1$. Then the STFT
covariance matrices are $\mathbf{R}_{n}[f]=\mathbf{a}_{n}[f]\mathbf{a}_{n}^{H}[f]$
for $n=1,\dots,N,$ where $\mathbf{a}_{n}[f]$ is a steering vector
with $a_{n,m}[f]=e^{j\Omega_{f}\tau_{n,m}}$ for $m=1,\dots,M$, $\Omega_{f}$
is the continuous-time frequency corresponding to frequency
index $f$, and $\tau_{n,m}$ is time delay of arrival for source $n$
at microphone $m$.

Now suppose that the positions of the microphones are randomly perturbed
so that $a_{n,m}[f]=e^{j\Omega_{f}(\tau_{n,m}+\Delta_{n,m})}$. If
$\Delta_{n,m}$ have independent Gaussian distributions with zero mean and
variance $\sigma^{2}$, then the off-diagonal elements of the ensemble
average covariance matrices are attenuated:
\begin{align}
\mathbf{\bar{R}}_{n,m_{1},m_{2}}[f] & =\mathbb{E}\left[e^{j\Omega_{f}(\tau_{n,m_{1}}-\tau_{n,m_{2}}+\Delta_{n,m_{1}}-\Delta_{n,m_{2}})}\right]\\
 & =\mathbf{R}_{n,m_{1},m_{2}}[f]\mathbb{E}\left[e^{j\Omega_{f}(\Delta_{n,m_{1}}-\Delta_{n,m_{2}})}\right]\\
 & =\mathbf{R}_{n,m_{1},m_{2}}e^{-\Omega_{f}^{2}\sigma^{2}},
\end{align}
where the last step comes from the moment-generating function. Because all off-diagonal elements are scaled
equally, we have
\begin{equation}
\bar{\mathbf{R}}_{n}[f]=e^{-\Omega_{f}^{2}\sigma^{2}}\mathbf{R}_{n}[f]+(1-e^{-\Omega_{f}^{2}\sigma^{2}})\boldsymbol{I}.\label{eq:cov_offsets}
\end{equation}

Substituting (\ref{eq:cov_offsets}) into (\ref{eq:div_gaussian})
and applying the Sherman-Morrison formula, it can be shown that the
Gaussian divergence between two source covariance matrices with these
Gaussian random offsets is
\begin{equation}
D(\bar{\mathbf{R}}_{1}[f],\bar{\mathbf{R}}_{2}[f])=\frac{M^{2}-\left|\mathbf{a}_{1}^{H}[f]\mathbf{a}_{2}[f]\right|^{2}}{2(e^{\Omega_{f}^{2}\sigma^{2}}-1)(e^{\Omega_{f}^{2}\sigma^{2}}-1+M)}.\label{eq:div_offsets}
\end{equation}
From this expression, the second-order statistics of the two sources
become more similar to each other as their unperturbed steering vectors
become closer together, as the uncertainty due to motion increases,
and as the frequency increases. Motion should have little impact if
$\Omega_{f}\sigma$ is small, that is, if the scale of the motion
is small compared to a wavelength. At high audible frequencies,
where acoustic wavelengths might be just a few centimeters, deformable
arrays will be quite sensitive to motion.

\subsection{Experimental measurements}

\begin{figure}
\begin{centering}
\includegraphics{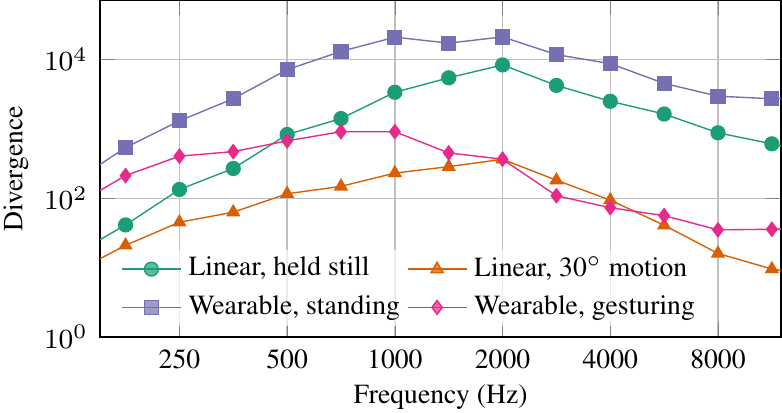}
\par\end{centering}
\caption{\label{fig:div-ensemble}Average divergence between source covariance
matrices.}
\end{figure}

The derivation above assumed independent motion
of all microphones. To confirm the predicted
trends\textemdash that spatial diversity decreases with frequency
and with amount of deformation\textemdash for real arrays with
more complex deformation patterns, the second-order statistics of
several deformable arrays were measured. Sample STFT covariance matrices were computed using
20-second pseudorandom noise signals produced sequentially by $N=5$
loudspeakers about 45\degree{}  apart in a half-circle around arrays of $M=12$ omnidirectional lavalier microphones.
One set of experiments used a linear array of microphones hanging
on cables from a pole that was manually rotated in a horizontal
plane. The hanging microphones swung by several millimeters relative to each other as they were moved. A second array used microphones affixed to
a hat and near the ears, chest, shoulders, and elbows
of a human subject who moved in different patterns. The arrays are shown in Figure \ref{fig:setup}.

\begin{figure}
\begin{centering}
\includegraphics{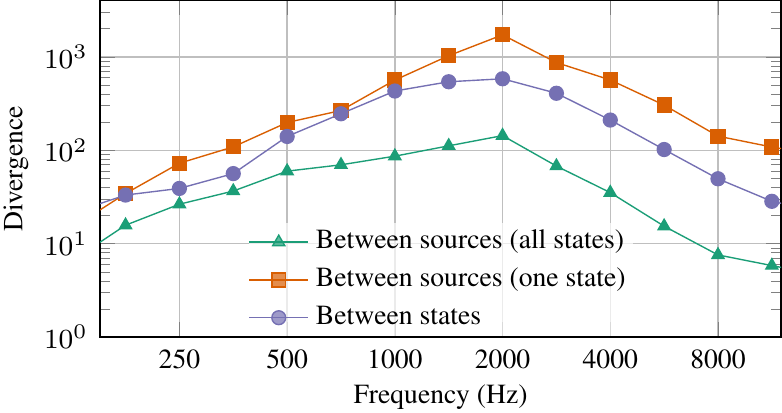}
\par\end{centering}
\caption{\label{fig:div-states}Divergence between sources and states for the hanging linear array. The between-source curves are the average divergence of the outer four sources with respect to the central source. The between-state curve is for the central source with the array at opposite ends of its range of motion, about 90\degree{} apart.}
\end{figure}

Figure \ref{fig:div-ensemble} shows the mean Gaussian divergence
between the long-term average STFT covariance matrices of the
central source and the four other sources for different array and
motion types. The nonmoving wearable array provides the greatest spatial
diversity between sources. The moving linear array provides the least.
For both arrays, motion causes the greatest penalty at higher frequencies,
as predicted.

With large deformations, it is difficult to distinguish the two sources
based on their long-term average statistics and it would be helpful
to use a time-varying model. Figure \ref{fig:div-states} shows the
divergence between ensemble average covariances of two sources over all states, $D(\bar{\mathbf{R}}_1[f],\bar{\mathbf{R}}_2[f])$; the divergence between their covariances in a single state,
$D(\tilde{\mathbf{R}}_{1}[f;\theta_{1}],\tilde{\mathbf{R}}_{2}[f;\theta_{1}])$; and the divergence
between two different states for the same source, $D(\text{\ensuremath{\tilde{\mathbf{R}}}}_{1}[f;\theta_{1}],\tilde{\mathbf{R}}_{1}[f;\theta_{2}]).$
At high frequencies, the two states are more different from each other
than the two sources are on average, suggesting that the ensemble covariance would not be useful for separation. The divergence between sources is an order
of magnitude larger within a single state than in the ensemble average.

\section{Static and Dynamic Beamforming}

\begin{figure}
\begin{centering}
\includegraphics{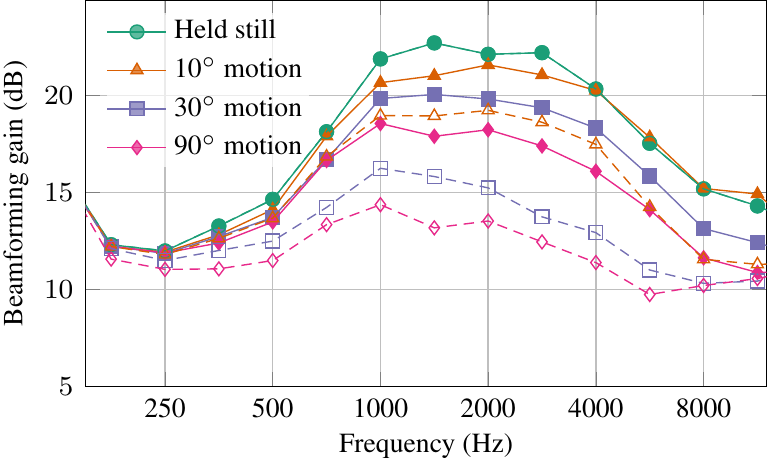}
\par\end{centering}
\caption{\label{fig:bf_pole}Beamforming performance with a linear array of
dangling microphones. Solid curves show dynamic beamforming
and dashed curves show static beamforming.}
\end{figure}

To demonstrate the impact of deformation on audio enhancement, the two arrays were used to separate
mixtures of speech sources using static and dynamic beamformers.
For each experiment, the STFT covariance matrices were estimated using
20 seconds of pseudorandom noise played sequentially from each loudspeaker while the array was moved.
The source signals are five 20-second anechoic speech
clips from different talkers in the VCTK corpus \cite{Veaux2017}.
The motion patterns produced by the human subject were similar but
not identical between the training and test signals.

Speech enhancement performance is measured using the mean improvement
in squared error between the input and output:
\begin{equation}
\mathrm{Gain}[f]=\frac{1}{N}\sum_{n=1}^{N}10\log10\frac{\sum_{t}\left|X_{1}[t,f]-D_{n}[t,f]\right|^{2}}{\sum_{t}\left|Y_{n}[t,f]-D_{n}[t,f]\right|^{2}}.
\end{equation}
Normally, the ground truth signals $D_{n}[t,f]$ could be measured
by recording each source signal in isolation. However, because the
motion patterns cannot be exactly reproduced between experiments,
it is impossible to know the ground truth signals received by
a moving array. To provide quantitative
performance measurements, the deformable arrays were supplemented
by a nonmoving microphone used as the reference $(m=1)$. To qualitatively evaluate a fully deformable array, the 
wearable-array experiments were repeated without the fixed microphone using the two
microphones near the ears as references; audio clips of these binaural
beamformer outputs are available on the first author's website\footnote{ryanmcorey.com/demos}.

\subsection{Dynamic beamforming with a linear array}

The rotating linear array is well suited to dynamic beamforming because
its state can be roughly described by its angle of rotation, which is easily
measured using near-ultrasonic pilot signals. In this experiment, the states formed a discrete set of about ten positions. Note that there is still some uncertainty within each state because the microphones are allowed to swing freely. Figure \ref{fig:bf_pole}
shows the average beamforming gain achieved by the linear array with different ranges of motion. Even small motion
from being held steady in the experimenter's hand causes poor high-frequency
performance. With 10\degree{}  rotation, the static beamformer performs a few decibels worse than the dynamic
motion-tracking beamformer. Dynamic beamforming is necessary for large
motion because the angle of rotation is larger than the angular spacing
between sources.

\subsection{Static beamforming with a wearable array}

\begin{figure}
\begin{centering}
\includegraphics{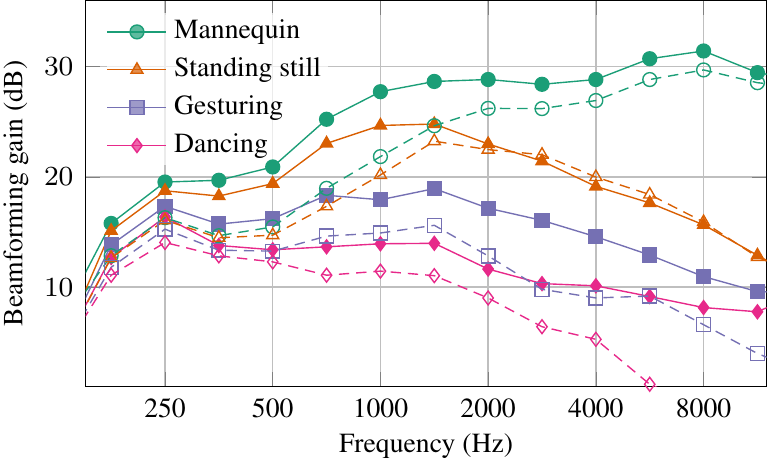}
\par\end{centering}
\caption{\label{fig:bf_human}Beamforming performance with a wearable microphone
array. Solid curves show full-rank beamformers dashed curves show rank-one beamformers.}
\end{figure}

The wearable array is more difficult to track dynamically because
there are many degrees of freedom in human motion. Figure \ref{fig:bf_human}
compares the performance of two static beamformers: one designed from
the full-rank average covariance matrix, and one designed
using a rank-one covariance matrix, that is, using an acoustic transfer
function measured from the training signals. 
For comparison with a truly nonmoving subject, the microphones
were placed on a plastic mannequin in the same configuration as on
the human subject. This motionless array performed well at the highest
tested frequencies. The human subject, even when trying to stand still, moved enough to destroy the phase coherence between
microphones at several kilohertz. These results suggest that researchers
should use caution when testing arrays on mannequins because
high-frequency performance might be different with live humans.

The full-rank covariance model outperforms the rank-one model even
for the motionless array at low frequencies. It improves robustness against both motion and diffuse
background noise. When the subject is gesturing---turning his head, nodding, and lifting and lowering his arms---or dancing in place by moving his arms, head, and torso, the full-rank beamformer outperforms the rank-one beamformer
by several decibels at all frequencies. However, at the highest tested
frequencies, the moving-array beamformers perform little better than
a single-channel Wiener filter, which would provide about 8 dB gain
for this five-source mixture.

\section{Conclusions}

The results presented here suggest that
deformable microphone arrays perform poorly at high frequencies. The
full-rank spatial covariance model can improve performance by several
decibels compared to a rank-one model, and dynamic beamforming that
tracks the state of the array provides even greater benefit. Even
so, it seems that deformable microphone arrays, including wearables, are most useful at low and mid-range frequencies. Fortunately, these are the frequencies most important for speech perception.

Deformable arrays are advantageous because they
can spread microphones across multiple devices or body parts. Thus, an 
array might combine rigidly-connected, closely-spaced
microphones for high frequencies with deformable, widely-spaced microphones
for low frequencies. Furthermore, as shown in \cite{corey2018async},
the full-rank covariance model can be used in
nonlinear, time-varying methods that aggregate
data from multiple wearable arrays. Large deformable arrays
can provide greater spatial diversity than small rigid arrays and
could be an important tool in spatial sound capture applications.

\bibliographystyle{ieeetr}
\bibliography{../master_references}

\end{document}